\documentclass{raa}            

\usepackage{graphicx,times}             
\usepackage{amsmath}

\begin{document}

   \title{An improved solution to geometric distortion using an orthogonal method
}

   \volnopage{Vol.0 (201x) No.0, 000--000}      
   \setcounter{page}{1}          

   \author{Huan-Wen Peng
      \inst{1,2,3,4}
   \and Qing-Yu Peng
      \inst{2,4}
   \and Na Wang
      \inst{2,4}
   }

   \institute{Yunnan Observatories, Chinese Academy of Sciences, Kunming 650216, China \\
        \and
            Department of Computer Science, Jinan University, Guangzhou 510632, China;{\it ~tpengqy@jnu.edu.cn}\\
        \and
            University of Chinese Academy of Sciences, Beijing 100049, China\\
        \and
            Sino-French Joint Laboratory for Astrometry, Dynamics and Space Science, Jinan University, Guangzhou 510632, China}

   \date{Received~~2016~~11~~03; accepted~~2016~~12~~17}

\abstract{The geometric distortion of CCD field of view has direct influence on the positional measurements of CCD observations. In order to obtain high precision astrometric results, the geometric distortion should be derived and corrected precisely. As presented in our previous work Peng et al. (2012), a convenient solution has been carried out and also been made with successful application to Phoebe's observations. In order to further improve the solution, an orthogonal method based on the Zernike polynomials is used in this work. Four nights of CCD observations including Himalia, the sixth satellite of Jupiter, and open clusters (NGC1664 or NGC2324) on each night have been processed to make an application. The observations were obtained from the 2.4 m telescope administered by Yunnan Observatories. The catalog UCAC4 was used to match reference stars in all of the CCD frames. The ephemeris of Himalia is retrieved from the Institut de M\'{e}canique C\'{e}leste et de Calcul des \'{E}ph\'{e}m\'{e}rides (IMCCE). Our results show that the means of observed minus computed (O-C) positional residuals are -0.034 and -0.026 arcsec in right ascension and declination, respectively. The corresponding standard deviations are 0.031 and 0.028 arcsec. The measurement dispersion is significantly improved than that by using our previous solution.
}

   \authorrunning{Huan-Wen Peng, Qing-Yu Peng \& Na Wang}            
   \titlerunning{An improved solution to geometric distortion}  

   \maketitle

%
%
\section{Introduction}           
\label{sect:intro}

The geometric distortion (called GD hereafter) which exists in both the space telescopes and ground-based telescopes has direct influence on astrometric precision of CCD observations. Gilmozzi et al. (\cite{Gilmozzi1995}) have found significant GD effects in the WFPC1 and WFPC2 of Hubble Space Telescope (HST). A very small field of view of 80$''$$\times$80$''$ for each CCD chip of WFPC2 has a maximum GD of about 5 pixels at the edge of its field (Anderson \& King~\cite{Anderson2003}). The astrometric potential of HST was just developed out after deriving the GD patterns and correcting its effects on positional measurements of planetary satellites (French et al.~\cite{French2006}). Anderson et al. (\cite{Anderson2006}) also applied the GD solution from HST to the ground-based 2.2 m telescope of ESO, and achieved a precision of $\sim$7 mas. In our previous works (Peng \& Fan~\cite{Peng2010}; Peng \& Tu~\cite{Peng2011}; Zhang et al.~\cite{Zhang2012}), GD effects of the 2.4 m and 1 m telescopes administered by Yunnan Observatories were first studied. As presented in Peng et al. (\cite{Peng2012}), an alternative GD solution which is different from the solution of Anderson \& King (\cite{Anderson2003}) has been carried out and also been made with successful application to Phoebe's observations. Since then, we have made several works with the new GD solution (Yang et al.~\cite{Yang2013}; Peng et al.~\cite{Peng2015}; Wang et al.~\cite{Wang2015}; Peng et al.~\cite{Peng2016}).

As presented in Peng et al. (\cite{Peng2012}), a dense star field should be observed in an overlapping scheme for deriving the GD patterns. As a practice, we may take multiple dithered exposures of the same sky field at different offsets in a pattern of ``+" (Anderson et al.~\cite{Anderson2006}) or ``\#" (Bellini \& Bedin~\cite{Bellini2010}). The offsets between any two neighboring CCD frames are about 1 arcmin in right ascension or in declination. In this way, a same star would appear in different overlapped CCD frames at different pixel positions for many times. According to the illustration showed in Peng et al.~(\cite{Peng2012}), an iteration method is used for deriving the GD patterns. In each iterative step, GDs of all the star images at different pixel positions could be obtained. Then all the GDs could be divided into many equal-area boxes, such as 19$\times$19 for the 2.4 m telescope. The average in each box would be indicative of the GD at its center if a gradual variation is assumed for the GD distributions. However, the scheme of dividing CCD field of view into many equal-area boxes is in some degree subjected to the distribution of star images. The GDs at the centers of some boxes can't be obtained when no star image exists in these areas.

As such, we try to adopt an orthogonal method presented in Plewa et al. (\cite{Plewa2015}). A list of twenty orthonormal basis vector fields which are based on the Zernike polynomials were used. For a detailed derivation, one can see Zhao \& Burge (\cite{Zhao2007,Zhao2008}). As showed in Plewa et al. (\cite{Plewa2015}), the radio source and massive black hole Sgr A* at the Galactic Center can be placed in the origin of an infrared astrometric reference frame with a precision of $\sim$ 0.17 mas in position (in 2009) and $\sim$ 0.07 mas yr$^{-1}$ in velocity, after correcting optical distortion in their NACO imager. This precision is a factor of 5 better than the previous results. This orthogonal method is used in this work to improve our previous GD solution. Specifically, instead of dividing the CCD field of view into many equal-area boxes, GDs of all the star images at different pixel positions in each iteration step were directly fitted by this group of basis vector fields. This method doesn't depend on the distribution of star images.

The contents of this paper are arranged as follows: in Section 2, the CCD observations are described; Section 3 presents the details of deriving GD patterns using the orthogonal method; in Section 4, we show the results and make discussions; and finally, in Section 5, conclusions are drawn.


\section{CCD Observations}
\label{sect:observation}

In order to analyze the improvements which the orthogonal method can obtain, four nights of CCD observations targeting Himalia, the sixth satellite of Jupiter, and open clusters (NGC1664 or NGC2324) were processed. These observations were obtained from the 2.4 m telescope (Fan et al.~\cite{Fan2015}) administered by Yunnan Observatories (IAU code O44, longitude E 100$^\circ$1$'$51$''$, latitude N 26$^\circ$42$'$32$''$, height 3193 m above sea level). The CCD detector used is the Yunnan Faint Object Spectrograph and Camera (YFOSC) instrument. Specifications of the 2.4 m telescope and YFOSC are listed in Table 1. Table 2 lists distributions of the CCD observations with respect to the observational dates. The observational dates were chosen according to the epoch when Jupiter was near its opposition. A total of 75 CCD frames of Himalia were obtained, as well as 176 CCD frames of calibration fields which were used for deriving GD patterns. The exposure time for each CCD frame is from 20s to 40s, depending on the meteorological conditions.

\begin{table}[htb]
\centering
\caption{Specifications of the 2.4 m telescope administered by Yunnan Observatories and the corresponding CCD detector.}
  \label{Tab1}
  \begin{tabular}{rr}
  \hline
  Parameters                           & 2.4 m telescope          \\
  \hline
  Approximate focal length             & 1920cm                   \\
  F-Ratio                              & 8                        \\
  Diameter of primary mirror           & 240cm                    \\
  CCD field of view(effective)         & 9$'\times$9$'$           \\
  Size of CCD array(effective)         & 1900$\times$1900         \\
  Size of pixel                        & 13.5$\mu m\times$13.5$\mu m$ \\
  Approximate scale factor             & 0.286$''$$/$pixel        \\
  \hline
\end{tabular}
\end{table}

\begin{table}[htb]
\begin{center}
\caption{CCD observations of Himalia and calibration fields by using the 2.4 m telescope administered by Yunnan Observatories. Column 1 shows the observational dates. Column 2 lists the open clusters observed. Column 3 and Column 4 list the numbers of observations for open clusters and Himalia, respectively. The Johnson-I filter was used in all observations.}
  \label{Tab2}
  \begin{tabular}{cccc}
  \hline\noalign{\smallskip}
  Obs dates        & Calibration fields &                & Himalia         \\
                   & Open clusters      & No.            & No.  \\
  \hline
  2015-02-07       & NGC2324            & 44             & 25             \\
  2015-02-08       & NGC2324            & 44             & 14            \\
  2015-02-09       & NGC2324            & 44             & 18             \\
  2015-02-10       & NGC1664            & 44             & 18             \\
  \hline
  Total            &                    & 176            & 75             \\
  \hline
\end{tabular}
\end{center}
\end{table}

\section{Details of Deriving GD patterns}

As presented in Peng et al.~(\cite{Peng2012}), an important relationship between the distortions at two different pixel positions for a common star can be derived, if the star was observed in two different CCD frames. The GDs in two CCD frames can be expressed as follows if the measured errors are temporarily neglected,
\begin{subequations}
\begin{equation}
dx_i=\Delta x_i-\frac{\hat{e}_icosD_i}{\hat{e}_jcosD_j}\Delta x_j +\frac{\hat{e}_icosD_i}{\hat{e}_jcosD_j}dx_j,
\end{equation}
\begin{equation}
dy_i=\Delta y_i-\frac{\hat{e}_i}{\hat{e}_j}\Delta y_j +\frac{\hat{e}_i}{\hat{e}_j}dy_j.
\end{equation}
\end{subequations}

In equations (1a) and (1b), all quantities with the suffix $i$ are associated with the $i$th CCD frame and the suffix $j$ with the $j$th CCD frame. $\hat{e}=cos\varphi/\rho$ is one of the estimated parameters in four-parameter linear transformation. $\rho$ and $\varphi$ are the approximate angular extent per pixel and the orientation of CCD chip used. $\Delta x$ and $\Delta y$ are the differences between the measured pixel location ($x_o$, $y_o$) of a star and the indirectly computed one ($x_c$, $y_c$) of the same star by using the four-parameter linear transformation with estimated parameters. The four parameters can be solved by a least-squares fitting. $D$ is declination of the tangent point on tangent plane of the celestial sphere for each CCD frame.

For a definite star, equation (1a) and (1b) can be solved if the star appears in $N$ ($N\gg2 $) CCD frames with different offsets. Then the distortions ($dx_i$, $dy_i$) of the star at different pixel positions in many CCD frames can be obtained. Furthermore, for all stars, the distortions at different pixel positions in all CCD frames can be collected. These distortions are divided into many equal-area boxes, such as 19$\times$19 for the 2.4 m telescope. The average in each box will be indicative of the GD at its center. Then the distortions of all star images at their pixel positions can be calculated through bilinear interpolation. For more details, one can see Peng et al. (\cite{Peng2012}).

As mentioned above, the scheme of dividing CCD field of view into many equal-area boxes is subjected to the distribution of star images. The GDs at the centers of some boxes which have no star images cannot be obtained. Thus we take use of an orthogonal method proposed in Plewa et al. (\cite{Plewa2015}) which does not depend on the distribution of star images. As analyzed in Plewa et al. (\cite{Plewa2015}), twenty orthonormal basis vector fields are needed to fully capture the spatial variability of the image distortion. These basis vector fields are derived based on the Zernike polynomials. For a detailed derivation, one can see Zhao \& Burge (\cite{Zhao2007,Zhao2008}). The explicit form of the twenty vector fields are listed in Table 3.

\begin{table}
\centering
\caption{Explicit form of the distortion model in terms of its basis vector fields. For a derivation, one can see Zhao \& Burge (2007, 2008). Column 1 shows the designation of each vector field. Column 2 lists the scale factor which should be multiplied by each vector field. The $d$ parameter in the scale factor represents the number of pixels in each dimension after that the original image pixel array is rescaled. Column 3 and column 4 list the components in two dimensions, respectively.}
  \label{Tab3}
  \begin{tabular}{llll}
  \hline
  $G(x,y)$ & Scale factor                              & $G_x(x,y)$                       & $G_y(x,y)$ \\
  \hline
  $S_2$    & $a_2=1$                                   & 1                                & 0 \\
  $S_3$    & $a_3=1$                                   & 0                                & 1 \\
  $S_4$    & $a_4=\sqrt{3}/d$                          & $\sqrt{2}x$                      & $\sqrt{2}y$ \\
  $S_5$    & $a_5=\sqrt{3}/d$                          & $\sqrt{2}y$                      & $\sqrt{2}x$ \\
  $S_6$    & $a_6=\sqrt{3}/d$                          & $\sqrt{2}x$                      & $-\sqrt{2}y$ \\
  $S_7$    & $a_7=\sqrt{24/(7d^4-24d^2+36)}$           & $\sqrt{6}xy$                     & $\sqrt{\frac{3}{2}}(x^2+3y^2-1)$ \\
  $S_8$    & $a_8=\sqrt{24/(7d^4-24d^2+36)}$           & $\sqrt{\frac{3}{2}}(3x^2+y^2-1)$ & $\sqrt{6}xy$ \\
  $S_9$    & $a_9=\sqrt{60/7d^4}$                      & $2\sqrt{3}xy$                    & $\sqrt{3}(x+y)(x-y)$ \\
  $S_{10}$ & $a_{10}=\sqrt{60/7d^4}$                   & $\sqrt{3}(x+y)(x-y)$             & $-2\sqrt{3}xy$ \\
  $S_{11}$ & $a_{11}=\sqrt{210/(81d^6-392d^4+560d^2)}$ & $2x(3x^2+3y^2-2)$                & $2y(3x^2+3y^2-2)$ \\
  $S_{12}$ & $a_{12}=\sqrt{105/(15d^6-84d^4+140d^2)}$  & $2\sqrt{2}x(2x^2-1)$             & $2\sqrt{2}y(1-2y^2)$ \\
  $S_{13}$ & $a_{13}=\sqrt{105/(30d^6-112d^4+140d^2)}$ & $2\sqrt{2}y(3x^2+y^2-1)$         & $2\sqrt{2}x(x^2+3y^2-1)$ \\
  $S_{14}$ & $a_{14}=\sqrt{70/3d^6}$                   & $2(x^3-3xy^2)$                   & $2(y^3-3x^2y)$ \\
  $S_{15}$ & $a_{15}=\sqrt{70/3d^6}$                   & $-2(y^3-3x^2y)$                  & $2(x^3-3xy^2)$ \\
  $T_4$    & $b_4=\sqrt{3}/d$                          & $\sqrt{2}y$                      & $-\sqrt{2}x$ \\
  $T_7$    & $b_7=\sqrt{24/(7d^4-24d^2+36)}$           & $\sqrt{\frac{3}{2}}(x^2+3y^2-1)$ & $-\sqrt{6}xy$ \\
  $T_8$    & $b_8=\sqrt{24/(7d^4-24d^2+36)}$           & $\sqrt{6}xy$                     & $-\sqrt{\frac{3}{2}}(3x^2+y^2-1)$ \\
  $T_{11}$ & $b_{11}=\sqrt{210/(81d^6-392d^4+560d^2)}$ & $2y(3x^2+3y^2-2)$                & $-2x(3x^2+3y^2-2)$ \\
  $T_{12}$ & $b_{12}=\sqrt{105/(15d^6-84d^4+140d^2)}$  & $2\sqrt{2}y(1-2y^2)$             & $2\sqrt{2}x(1-2x^2)$ \\
  $T_{13}$ & $b_{13}=\sqrt{105/(30d^6-112d^4+140d^2)}$ & $2\sqrt{2}x(x^2+3y^2-1)$         & $-2\sqrt{2}y(3x^2+y^2-1)$ \\
  \hline
\end{tabular}
\end{table}

As showed in Table 3, in order to apply these twenty basis vector fields in our previous GD solution, there are several steps to be accomplished. Firstly, the pixel positions of star images are rescaled that the pixel coordinates become much more smaller than the original ones. In such a way, the numerical computations can be more precise. Secondly, the scale factors listed in Table 3 can be calculated according to the orthonormality for any two vector fields. Thirdly, a least-squares fitting is applied for deriving the coefficients of each vector field. Finally, the distortion at any pixel position can be directly calculated by using the vector field of GD which is solved in the previous step.

Specifically, as illustrated in Zhao \& Burge (\cite{Zhao2007,Zhao2008}), the two components of each vector field which are $G_x(x,y)$ and $G_y(x,y)$ listed in Table 3 are defined over a unit circle. However, CCD chips are always square or rectangle. Thus the transformation from an unit circle to a square or a rectangle must be applied. In practice, if $\vec{B}$ and $\vec{C}$ are two vector fields defined over an unit circle, we define their inner product as
\begin{equation}
\left(\vec{B},\vec{C}\right)=\frac{1}{\pi}\iint\left(\vec{B}\cdot\vec{C}\right)dxdy,
\end{equation}
where $\pi$ is the area of an unit circle. Then the inner product of two vector fields which are $\vec{G}_i$ and $\vec{G}_j$ ($i,j=1\sim20$) defined over a square or a rectangle is
\begin{equation}
\left(\vec{G}_i,\vec{G}_j\right)=\frac{1}{A}\iint\left(\vec{G}_i\cdot\vec{G}_j\right)dxdy,
\end{equation}
where $A$ is the area of a square or a rectangle. In order to satisfy the orthonormality, the inner product of any two vector fields defined over a square or a rectangle is
\begin{equation}
\left(\vec{G}_i,\vec{G}_j\right)=\delta_{ij}=
\begin{cases}
1,\mbox{ if }i=j \\
0,\mbox{ if }i\ne j.
\end{cases}
\end{equation}

According to equation (4), the scale factors listed in Table 3 can be calculated. The values of scale factors depend on the rescaled size of image pixel array. In practice, the scale factors according to a square pixel array which has $d$ pixels in each dimension are listed in Table 3. An iterative method is used for deriving the vector field of GD. Specifically, in a definite iteration step, the GDs of all star images are fitted by the distortion model in Table 3. The GD pattern in this iteration step is added to the final GD pattern. Then the GDs of all star images in the next iteration step are solved again after GD corrections are made. When the values of GD pattern in an iteration step are within 0.01 pixel, the iteration process is stopped. After the final vector field of GD is solved, the distortions of all star images at their pixel positions can be directly computed. Finally, the GD corrections can be applied.

\section{Results and Discussions}
\label{sect:results}

\begin{figure}[tb]
\centering
\includegraphics[width=1.02\linewidth]{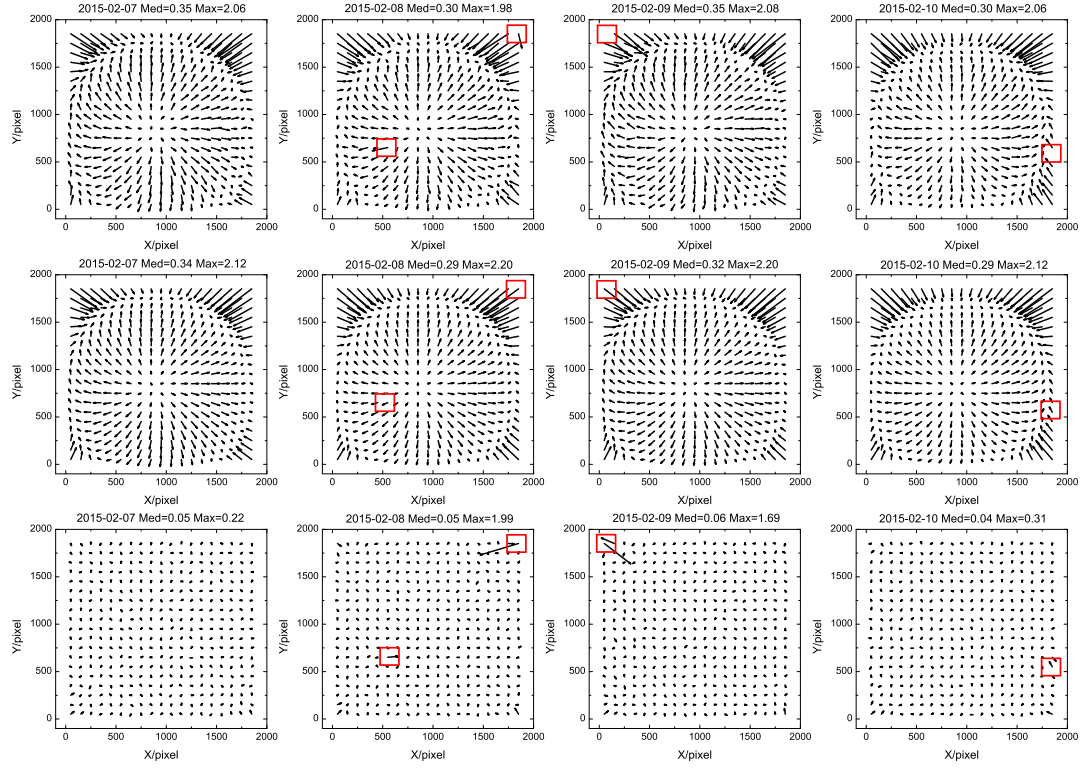}
\caption{GD patterns for the 2.4 m telescope administered by Yunnan Observatories.
The upper four panels in the first row show the GD patterns derived by previous solution. The four panels in the second row show the GD patterns derived by improved solution. The lower four panels in the third row show the differences between GD patterns in the second row and first row. All observations were made with a filter-I. In each panel, the observational date, the median and maximum GD values are listed on the top in units of pixels. A factor of 200 is used to exaggerate the magnitude of each GD vector.}
\label{Fig1}
\end{figure}

\begin{figure}[tb]
\centering
\includegraphics[width=1.0\linewidth]{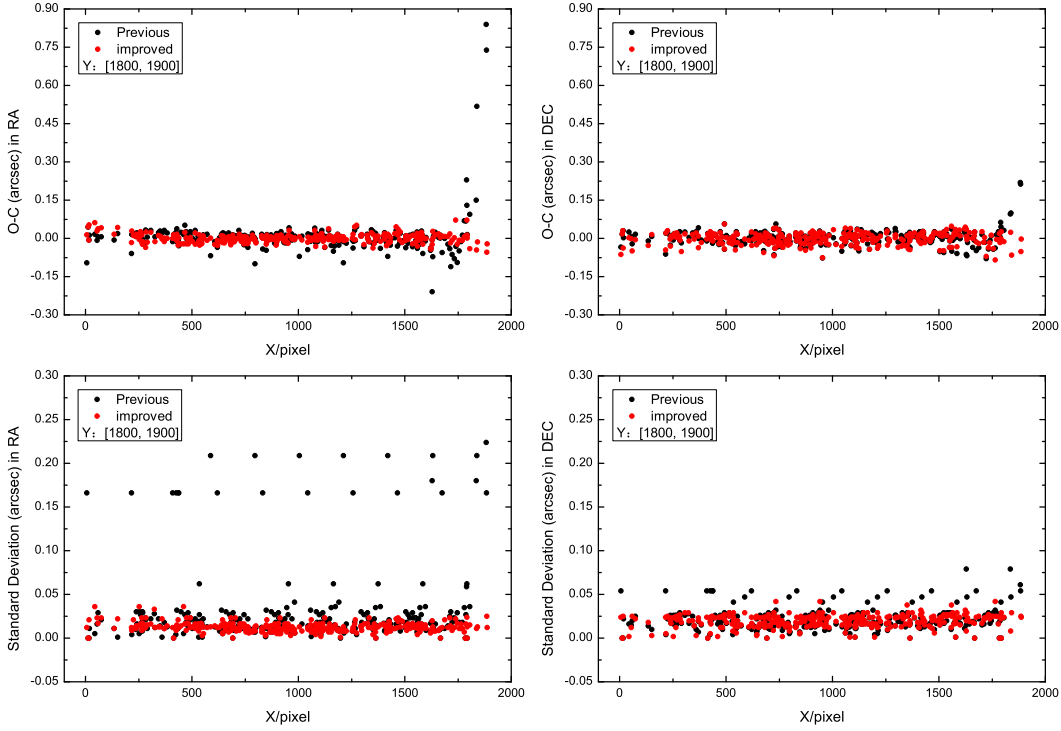}
\caption{(O-C) residuals of common stars and their standard deviations (SDs) on February 8. The selected pixel area has the range of coordinate x from 0 to 1900 and the range of coordinate y from 1800 to 1900. The upper two panels show the (O-C) residuals in right ascension and declination, respectively. The lower two panels show the SDs in each direction. The dark and red points represent the (O-C) residuals or the SDs of common stars for the previous and improved solution, respectively. Some dark points in the lower two panels have the same SD value because the same star appears at different pixel positions in several different CCD frames.}
\label{Fig2}
\end{figure}

The catalog UCAC4 (Zacharias et al.~\cite{Zacharias2013}) was chosen to match reference stars in all CCD frames. The minimum and maximum numbers of UCAC4 reference stars available for astrometric reduction of Himalia are 7 and 18, respectively. Observed positions are derived relative to these UCAC4 reference stars by using a plate model with four constants. However, this is accurate only after all the astrometric effects, including GD effects, are taken into account (Peng et al.~\cite{Peng2012}).

Fig. 1 shows the GD patterns derived by both the previous and improved solutions, and also differences between the GD patterns. One can see that the distributions and variations of GD vectors are more smooth after that the improved solution was used. From the first row of Fig. 1, we can see the inconsistent GD vectors. Specifically, the areas marked by red rectangles in the GD pattern on February 8 have GD vectors which are inconsistent with the nearby ones. Especially for the top right corner of the GD pattern on February 8, the amount of star images in this area is only five. The magnitudes of these stars are between 15$\sim$17. Thus measured errors would be the primary source and make the GD values incorrect, especially for faint stars. The top left corner of the GD pattern on February 9 has no GD vector, because there are no star images in this area. The area marked by red rectangle in the GD pattern on February 10 has no GD vector either. However, the corresponding areas in the four GD patterns of second row by using the improved solution have reasonable GD vectors. We can clearly see these differences from the third row.

From the third row of Fig. 1 we can see that the GDs in most areas have only subtle differences between the GD patterns derived by the previous and improved solution. In order to show the improvements made by improved solution, the (O-C) residuals and standard deviations (SDs) of common stars falling into the marked area in the top right corner of GD pattern on February 8 are drawn in Fig. 2. The SD of one definite star is based on its (O-C) residuals in many different CCD frames. The selected rectangular pixel area has the range of coordinate x from 0 to 1900 and the range of coordinate y from 1800 to 1900. Fig. 2 shows the details. The (O-C) residuals and SDs of some stars in the top right corner of GD pattern on February 8 are significantly improved, because the wrong GD vector in the first row of Fig. 1 is reasonably calculated in the second row. These improvements give proof that the GD solution with the orthogonal method is more suitable for deriving GD patterns.

In order to check how much improvements on the positional precision could be obtained by using the GD solution with orthogonal method, four nights of CCD observations of Himalia were processed. The observed positions of Himalia were compared to the ephemerides retrieved from the IMCCE which include satellite ephemeris by Emelyanov (\cite{Emelyanov2005}) and planetary ephemeris INPOP13c (Fienga et al.~\cite{Fienga2015}). Fig. 3 shows the (O-C) residuals of positions of Himalia with respect to the observational epochs. Table 4 lists the statistics of (O-C) residuals of Himalia by using both the previous and improved GD solutions. We can see that the internal agreement or precision on February 8 has relatively high improvement than the other nights. The means of (O-C) residuals for all data after using improved GD solution are -0.034$''$ and -0.026$''$ in right ascension and declination, respectively. The corresponding standard deviations are 0.031$''$ and 0.028$''$.

\begin{figure}[htb]
\centering
\includegraphics[width=1.0\linewidth]{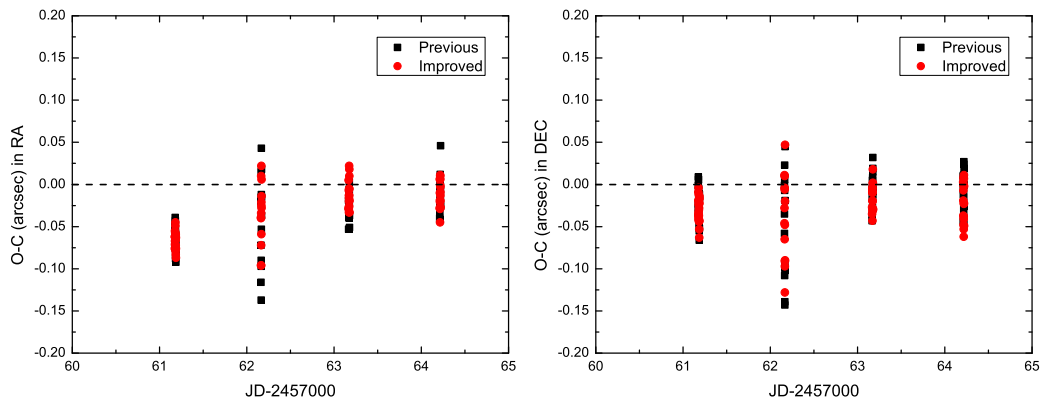}
\caption{(O-C) residuals of the topocentric apparent positions of Himalia compared to the ephemeris retrieved from the IMCCE which include satellite theory by Emelyanov (\cite{Emelyanov2005}) and planetary ephemeris INPOP13c, with respect to the Julian Dates. The dark and red points represent the (O-C) residuals by using previous and improved GD solutions, respectively.}
\label{Fig3}
\end{figure}

\begin{table}[htb]
\centering
\caption{Statistics of (O-C) residuals of the positions of Himalia by using both the previous and improved GD solutions. Column 1 shows the observational dates. Column 2 shows which GD solution was used. The following columns list the means of (O-C) residuals and their standard deviations (SDs) in right ascension and declination, respectively. All units are in arcseconds.}
  \label{Tab4}
  \begin{tabular}{rrrrrr}
  \hline
  Obs dates        & GD           &$\langle$O-C$\rangle$ & SD     &$\langle$O-C$\rangle$ & SD        \\
                   & Solution     & RA     &        & DEC    &           \\
  \hline
  2015-02-07       & Previous     & -0.065 & 0.014  & -0.023 & 0.018     \\
                   & Improved     & -0.066 & 0.010  & -0.028 & 0.014     \\
  2015-02-08       & Previous     & -0.042 & 0.054  & -0.052 & 0.062     \\
                   & Improved     & -0.035 & 0.036  & -0.040 & 0.050     \\
  2015-02-09       & Previous     & -0.029 & 0.016  & -0.006 & 0.018     \\
                   & Improved     & -0.012 & 0.017  & -0.016 & 0.016     \\
  2015-02-10       & Previous     & -0.011 & 0.021  & -0.009 & 0.024     \\
                   & Improved     & -0.011 & 0.016  & -0.021 & 0.024     \\
  \hline
  Total            & Previous     & -0.039 & 0.034  & -0.021 & 0.035     \\
                   & Improved     & -0.034 & 0.031  & -0.026 & 0.028     \\
  \hline
\end{tabular}
\end{table}

\section{Conclusions}
\label{sect:conclusion}

In this paper, we improve our previous GD solution by using an orthogonal method based on the Zernike polynomials. A total of 75 CCD observations obtained from the 2.4 m telescope administered by Yunnan Observatories were processed. The precision of astrometric position of Himalia is significantly better with the improved GD solution. The results show that means of (O-C) residuals of Himalia are -0.034$''$ and -0.026$''$ in right ascension and declination, respectively. The corresponding standard deviations are 0.031$''$ and 0.028$''$. As is well known, the new catalog Gaia DR1 (Gaia Collaboration et al.~\cite{GaiaCollaboration2016b}) was released on September 14, 2016, after that the Gaia space probe (Gaia Collaboration et al.~\cite{GaiaCollaboration2016a}) has been launched on December 19, 2013. This catalog represents a huge improvement in the available fundamental stellar data and practical definition of the optical reference frame (Lindegren et al.~\cite{Lindegren2016}). The unprecedent astrometric precision of reference stars can allow us to obtain quite higher positional precision of targets. Our improved GD solution is also useful for astrometric data reduction in the future.

\begin{acknowledgements}
We acknowledge the support of the staff at the Lijiang 2.4 m telescope. Funding for the telescope has been provided by CAS and the People¡¯s Government of Yunnan Province. This work is financially supported by the National Natural Science Foundation of China (grant nos. U1431227 and 11273014).
\end{acknowledgements}

%
%

\label{lastpage}

\end{document}